\documentclass[review]{elsarticle}

\usepackage{lineno,hyperref}
\usepackage{amsmath}
\journal{Journal of Crystal Growth}

\begin{document}

\begin{frontmatter}

\title{Growth and characterization of CaFe$_{1-x}$Co$_x$AsF single crystals by CaAs flux method}

\author[ShTech,SIMIT]{Yonghui Ma}
\author[SIMIT,ShU]{Kangkang Hu}
\author[SIMIT]{Qiucheng Ji}
\author[SIMIT]{Bo Gao}
\author[SIC]{Hui Zhang}
\author[SIMIT]{Gang Mu\corref{mycorrespondingauthor}}
\cortext[mycorrespondingauthor]{Corresponding author. Tel.:
86-21-62511070.} \ead{mugang@mail.sim.ac.cn}
\author[SIC]{Fuqiang Huang}
\author[ShTech,SIMIT]{Xiaoming Xie}

\address[ShTech]{School of Physical Science and Technology, ShanghaiTech University, Shanghai 201210, China}
\address[SIMIT]{State Key Laboratory of Functional Materials for Informatics and Shanghai Center for Superconductivity, Shanghai Institute of
Microsystem and Information Technology, Chinese Academy of Sciences,
Shanghai 200050, China}
\address[ShU]{College of Sciences, Shanghai University, Shanghai 200444, China}
\address[SIC]{CAS Key Laboratory of Materials for Energy Conversion, Shanghai Institute of Ceramics, Chinese Academy of Sciences, Shanghai
200050, China}

\begin{abstract}
Millimeter sized single crystals of CaFe$_{1-x}$Co$_x$AsF were grown
using a self-flux method. It is found that high-quality single
crystals can be grown from three approaches with different initial
raw materials. The chemical compositions and crystal structure were
characterized carefully. The c-axis lattice constant is suppressed
by the Co substitution. Superconductivity with the critical
transition $T_c$ as high as 21 K was confirmed by both the
resistivity and magnetic susceptibility measurements in the sample
with $x$ = 0.12. Moreover, it is found that $T_c$ can be enhanced
for about 1 K under the very small hydrostatic pressure of 0.22 GPa,
which is more quickly than that reported in the polycrystalline
samples. Our results is a promotion for the physical investigations
of 1111 phase iron-pnictide superconductors.
\end{abstract}


\end{frontmatter}


\section{Introduction}
Investigations on the physical properties of 1111 phase of the
iron-pnictide superconductors are restricted remarkably, compared
with the 122 phase and 11 phase, due to the difficulties in
obtaining high-quality and sizable single crystals. NaCl and KCl
were first used as the flux and small single crystals with small
sizes of 20-70 $\mu$m can be obtained~\cite{NaCl} because of the low
solubility of NaCl/KCl flux. Then the high pressure method was
developed and the crystal size was enhanced to several hundred
micrometers ~\cite{pressure,review}. After that, a new flux NaAs
used under ambient pressure was reported to improve the crystal
size~\cite{NaAs}, but the crystal quality is not sufficient for the
physical investigations possibly due to the extrinsic contaminants
induced by the flux. A noteworthy progress in this field was made by
our group by using the CaAs flux, where millimeter sized single
crystals of CaFeAsF with high quality was grown
successfully~\cite{CaAs}. It is the parent compound of the
fluorine-based 1111 phase of the iron-pnictide superconductors,
whose $T_c$ can be increased to above 50 K by doping rare-earth
elements to the sites of the alkaline-earth elements and to above 20
K by doping transition-metal elements to the sites of Fe
~\cite{SrFeAsF1,SrFeAsF2,CaFeAsF-Nd,CaFeAsF-Co,SrFeAsF-Sm,SrFeAsF-La}.

To grow the doped single crystals with superconductivity was found
to be more difficult than we expected. After many trials, single
crystals of Co-doped CaFeAsF were demonstrated to be the
breakthrough and grown successfully. In this paper, we report the
technical approaches to grow the CaFe$_{1-x}$Co$_x$AsF single
crystals with a high quality and large size. The compositions and
structure of the crystals were determined. The superconducting
properties were confirmed by the low-temperature resistivity and
magnetization measurements. We also checked the influence of the
pressure on the superconductivity.

\section{Experiment}

The CaFe$_{1-x}$Co$_{x}$AsF single crystals were grown on the basis
of the successful growth of the parent phase CaFeAsF, where a CaAs
self-flux method was developed~\cite{CaAs}. It was found that the
CaFeAsF single crystals can been grown successfully with different
initial raw materials as the following formulas (1) and (2)
describe.
\begin{equation}
CaAs+ 0.5 FeF_{2}+ 0.5 Fe + 4CaAs (Flux) \Rightarrow CaFeAsF + 4CaAs
(Flux)
\end{equation}
\begin{equation}
CaAs+ CaF_{{2}}+ Fe_{{2}}As + 8CaAs (Flux) \Rightarrow 2CaFeAsF +
8CaAs (Flux)
\end{equation}
Based on the formula (1), two approaches (see formula (3) and (4))
was tried to substitute Co to the site of Fe and both were found to
be successful in growing the single crystals.
\begin{equation}
\begin{split}
CaAs+0.5 FeF_{{2}}&+ (0.5-x) Fe+ x Co+ 4CaAs (Flux)\\
&\Rightarrow CaFe_{1-x}Co_{x}AsF + 4CaAs (Flux) \end{split}
\end{equation}
\begin{equation}\begin{split}
CaAs+(0.5-x)F&eF_{{2}}+x CoF_{{2}}+ 0.5 Fe + 4CaAs (Flux)\\
&\Rightarrow CaFe_{1-x}Co_{x}AsF + 4CaAs (Flux)
\end{split}\end{equation}
Starting from formula (2), only Fe$_2$As can be substituted by
Co$_2$As and we have the following formula to grow the single
crystals of CaFe$_{1-x}$Co$_{x}$AsF.
\begin{equation}\begin{split}
CaAs + CaF_{{2}} + (1&-x) Fe_{{2}}As+ x Co_{\rm{2}}As+ 8CaAs (Flux)\\
&\Rightarrow 2CaFe_{1-x}Co_{x}AsF + 8CaAs (Flux)
\end{split}\end{equation}

In practice, the arsenide precursors CaAs, Fe$_2$As and Co$_2$As
were first synthesized by heating a mixture of Ca granules (purity
99.5\%, Alfa Aesar), Fe powder (99+\%, Alfa Aesar), Co powder
(99.8\%, Alfa Aesar), and As grains (purity 99.995\%, Alfa Aesar) at
700-800 $^{\circ}$C for 10-15 h in an evacuated quartz tube. Then
homemade precursors and other reagents was mixed according to
formula (3), (4) and (5) and placed into a crucible. The purities of
the reagents are FeF$_2$ (purity 99\%, Alfa Aesar), CoF$_2$ (purity
98\%, Alfa Aesar), and CaF$_2$ (purity 99.95\%, Alfa Aesar). All the
weighing and mixing procedures were carried out in a glove box with
a protective argon atmosphere. Finally, the crucible was sealed in a
quartz tube with vacuum. The quartz tube was heated at 950
$^{\circ}$C for 40 h firstly, and then it was heated up to 1230
$^{\circ}$C where it remained for 20 h. Finally it was cooled down
to 900 $^{\circ}$C at a rate of 2 $^{\circ}$C/h and followed with a
quick cooling down to room temperature. In the end, by exposing the
resultant to air in the fume hood for a few days, the flux CaAs
decomposed and the single crystals appeared. We found that all of
the three methods are successful to obtain the
CaFe$_{1-x}$Co$_{x}$AsF single crystals and the quality of the
single crystals is very similar to each other with a small
fluctuation of $T_c$ for about $\pm0.5$ K, so we will not
distinguish them in our paper.

The microstructure was examined by scanning electron microscopy
(SEM, Zeiss Supra55). The composition of the single crystals was
checked and determined by energy dispersive x-ray spectroscopy (EDS)
measurements on an Bruker device with the model Quantax200. The
crystal structure and lattice constants of the materials were
examined by a DX-2700 type powder x-ray diffractometer using Cu
K$_\alpha$ radiation. The electrical resistivity was measured using
a four probe technique on the physical property measurement system
(Quantum Design, PPMS). The magnetic susceptibility measurement was
carried out on the magnetic property measurement system (Quantum
Design, MPMS 3) with the magnetic field oriented parallel to the
ab-plane of the samples. The magnetization susceptibility under
different pressures was obtained using the Mcell 10 (Almax easylab).

\section{Results and discussions}

\begin{figure}
\includegraphics[width=11cm]{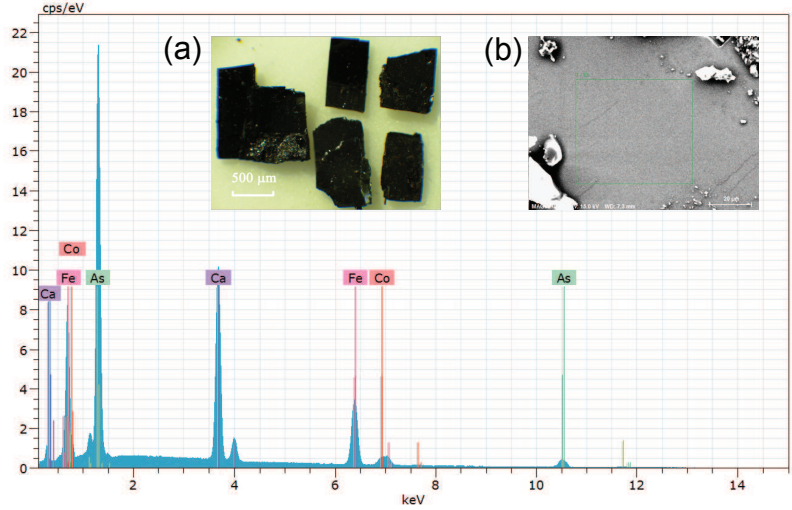}

\caption {(color online) The EDS microanalysis spectrum taken on the
crystal with the nominal doping level $x_{\mathrm{nominal}}$ = 0.15.
The inset (a) and (b) are the surface pictures of this crystal taken
with the optical microscope and SEM, respectively. } \label{fig1}
\end{figure}

\begin{table}
\centering \caption{Compositions of the crystal with
$x_{\mathrm{nominal}}$ = 0.15 characterized by EDS measurements.}
\begin{tabular}
{ccccccc}\hline \hline
Element &       Weight (\%)   &  Atomic (\%)  \\
\hline
Ca          & 23.43   & 31.91    \\
Fe          & 27.62   & 26.99   \\
Co          & 3.90    & 3.61   \\
As          & 43.30   & 31.54   \\
 \hline \hline
\end{tabular}
\label{tab.1}
\end{table}

The morphology of the single crystals was examined by the optical
microscope and the scanning electron microscopy, which are shown in
figures \ref{fig1}(a) and (b) respectively. The crystals are black
in color and show the flat surfaces. The typical crystal size was
found to be as large as 1 mm $\times$ 0.7 mm $\times$ 0.06 mm. Some
impurities on the surface of the sample are the remaining flux CaAs.
The composition of the crystals was examined by the EDS analysis and
one of the typical EDS spectrums for $x_{\mathrm{nominal}}$ = 0.15
is shown in figure \ref{fig1} and table \ref{tab.1}. The square
frame in figure 1(b) indicates the measuring field of EDS. Note that
the content of F is not precise as other elements because the EDS is
not sensitive to light element. The EDS data revealed that the ratio
of Ca:(Fe+Co):As is close to 1:1:1, and the actual composition of Co
can be determined to be about $x_{\mathrm{actual}}$ = 0.12.
Hereinafter the variable $x$ denotes the actual doping level
$x_{\mathrm{actual}}$ in this paper.

\begin{figure}
\includegraphics[width=11cm]{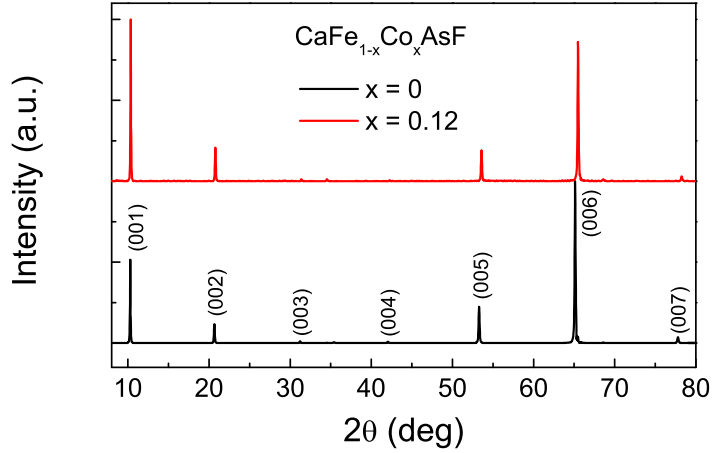}
\caption {Typical XRD patterns of CaFeAsF (black line) and
CaFe$_{1-x}$Co$_{x}$AsF with $x_{\mathrm{actual}}$ = 0.12 (red
line). } \label{fig2}
\end{figure}

The structure of the crystals was examined by a powder XRD
measurement, where the x-ray was incident on the ab-plane of the
sample. The XRD spectrum for the undoped crystal CaFeAsF is
displayed together with the sample with $x$=0.12 in figure
\ref{fig2} for a comparison. Only sharp peaks with the index (00
$l$) could be observed, suggesting a high c-axis orientation of the
crystals. It is clear that the diffraction peaks for the doped
sample shifts to the right sides compared with the parent compound,
suggesting the shrinkage of the crystal along the $c$ axis. The raw
data of XRD was analyzed by PowderX
software~\cite{powerX-1,powerX-2}, where the zero-shift, K$_
{\alpha2}$-elimination and other factors were taken into account.
The c-axis lattice parameters was calculated to be 8.584 \r{A} and
8.547 \r{A} for the samples with $x=0$ and $x=0.12$ respectively,
which is consistent with the results of the polycrystalline samples
~\cite{CaFeAsF-poly}.

\begin{figure}
\includegraphics[width=11cm]{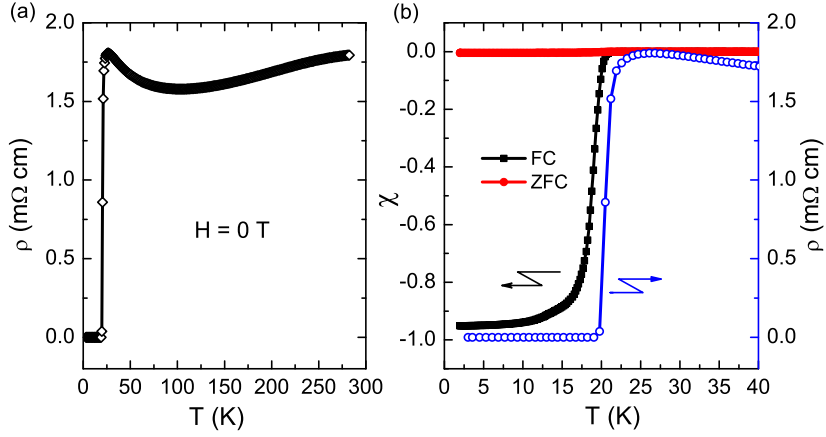}
\caption {(a) Temperature dependence of resistivity for the
CaFe$_{1-x}$Co$_{x}$AsF single crystal with $x_{\mathrm{actual}}$ =
0.12 measured in a wide temperature range 0 - 300 K under zero
magnetic field. (b) The normalized magnetic susceptibility measured
in zero-field-cooled (ZFC) and field cooled (FC) models and the
resistivity data in the low temperature range near the
superconducting transition. } \label{fig3}
\end{figure}

Temperature dependence of resistivity in the temperature range from
0 to 300 K for the single crystal CaFe$_{1-x}$Co$_{x}$AsF with $x$ =
0.12 is shown in figure \ref{fig3}(a). The $\rho-T$ curve is lowered
gradually with the decrease of temperature above 100 K, and followed
by a clear upwarp below 100 K, which suggests that the sample is
slightly underdoped. The onset of the superconducting transition
appears at about 22 K, whereas the zero resistivity is reached at
about 20 K. The dc magnetic susceptibility for the same sample was
measured under a magnetic field of 10 Oe in zero-field-cooling and
field-cooling processes, which is presented in figure \ref{fig3}(b)
with temperature between 0 and 40 K. In order to minimize the effect
of the demagnetization, the magnetic field was applied parallel to
the ab-plane of the crystal. The absolute value of magnetic
susceptibility $\chi$ is over 95\%, indicated a high superconducting
volume fraction of our sample. An enlarged $\rho-T$ curve is also
shown in figure \ref{fig3}(b) in order to be compared with
susceptibility curve expediently. The superconducting transition
temperature revealed by the $\rho-T$ curve is slightly higher than
that by the $\chi-T$ curve, which is rather reasonable and common
for the compound superconductors.

\begin{figure}
\includegraphics[width=11cm]{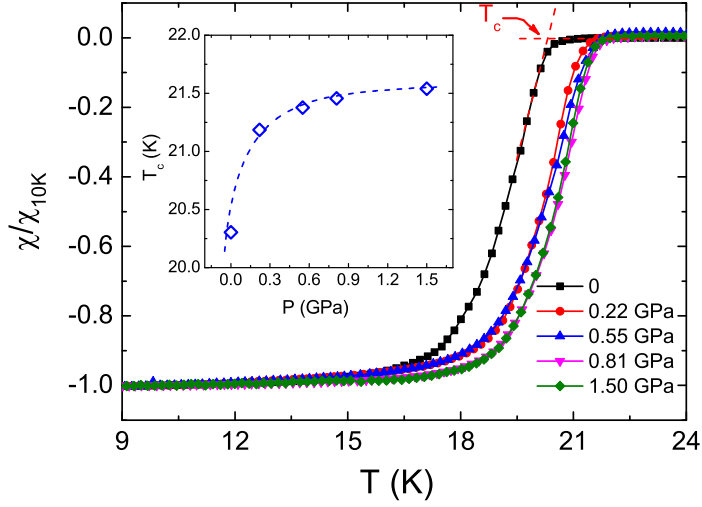}
\caption {Temperature dependence of the magnetic susceptibility
under hydrostatic pressures of up to 1.5 GPa for
$x_{\mathrm{actual}}$ = 0.12 below  24 K. The inset presents the
pressure dependence of $T_c$ for this sample. These data were
obtained using Mcell 10. } \label{fig4}
\end{figure}

We investigated the influence of pressure on the superconductivity
of our samples. The magnetization susceptibility under different
pressures for the sample CaFe$_{1-x}$Co$_{x}$AsF with $x$ = 0.12 was
obtained using Mcell 10. The pressure was generated using a Be-Cu
pressure cell with a Teflon cup which was filled with the
pressure-transmitting medium. The actual pressure was calibrated and
determined by the shift of the superconducting transition
temperature of pure Tin. The results of the $\chi-T$ data with
pressure varying from 0 to 1.5 GPa are presented in figure
\ref{fig4}. The intensity of the diamagnetic signals are variable
for different pressure due the change of the orientation of the
sample in the pressure cell. All of the susceptibility curves have
been normalized by the dada at 10 K for the convenience of
distinguishing their changes. It is clearly demonstrated that a
pressure as small as 0.22 GPa can enhance $T_c$ for about 1 K, and
then, with further increasing of the pressure, the change of $T_c$
becomes inconspicuous. The inset of figure \ref{fig4} shows the
pressure dependence of $T_c$ for the sample. With the increasing of
pressure, $T_c$ rises steeply initially and then it becomes smooth
and stable around a maximum value 21.5 K. Actually, the pressure
effect on the superconductivity of CaFe$_{1-x}$Co$_{x}$AsF has been
studied on the polycrystalline samples~\cite{CaFeAsF-pressure}. The
amplitude of the $T_c$ enhancement of our samples induced by
pressure is comparable with that observed in the polycrystalline
samples, which is much smaller than the LaFeAsO$_{1-x}$F$_x$.
Moreover, it is found that our observations are more similar to the
polycrystalline sample with $x=0.1$, because $T_c$ was reported to
decrease above merely 0.4 GPa for the samples with $x=0.15$ and 0.2.
However, $T_c$ increase more quickly for our single-crystalline
samples.

\section{Conclusions}

In this work, millimeter sized single crystals of
CaFe$_{1-x}$Co$_x$AsF were successfully grown by CaAs self-flux
method via three approaches with different initial raw materials.
The chemical compositions, crystal structure, resistivity and
magnetization susceptibility were investigated systematically. Both
the resistivity and magnetic susceptibility show that the $T_c$ of
CaFe$_{1-x}$Co$_{x}$AsF with $x$ = 0.12 is about 21 K, which can be
enhanced for about 1 K under the hydrostatic pressure of 0.22 GPa.
Moreover, it is found that $T_c$ of the present single-crystalline
samples increases more quickly with pressure compared with the
polycrystalline samples. Our results supply a platform to study the
intrinsic properties of the 1111 phase of iron-pnictide
superconductors.

\section*{Acknowledgments}
This work is supported by the Natural Science Foundation of China
(No. 11204338), the ``Strategic Priority Research Program (B)" of
the Chinese Academy of Sciences (No. XDB04040300 and XDB04030000)
and the Youth Innovation Promotion Association of the Chinese
Academy of Sciences (No. 2015187).

\section*{References}


\end{document}